\documentclass[journal,comsoc]{IEEEtran}
\usepackage[T1]{fontenc}% optional T1 font encoding

\ifCLASSINFOpdf
\else
\fi
\usepackage{amsmath}
\usepackage[cmintegrals]{newtxmath}

\usepackage{epsfig}
\usepackage{color}

\usepackage{verbatim}
\usepackage{graphicx}
\usepackage{float}
\usepackage{color}
\usepackage{fancyhdr}
\usepackage{authblk}
\usepackage{tikz-cd}
\usepackage{tikz}
\usepackage{amsmath} % assumes amsmath package installed
\usepackage[normalem]{ulem}
\newtheorem{definition}{Definition}[section]
\newtheorem{lemma}[definition]{Lemma}
\newtheorem{theorem}[definition]{Theorem}
\newtheorem{proposition}[definition]{Proposition}

\newtheorem{remarkth}[definition]{Remark}
\newtheorem{exampleth}[definition]{Example}
\newenvironment{remark}{\begin{remarkth}\upshape}{\hfill$\diamond$\end{remarkth}}
\newenvironment{example}{\begin{exampleth}\upshape}{\hfill$\diamond$\end{exampleth}}
\renewcommand{\emph}[1]{{\bfseries\itshape{#1}}}

\usepackage{amsfonts}
\usepackage{graphicx}
\usepackage{amsmath, amssymb}
\usepackage{mathrsfs}

\usepackage{multirow}
\usepackage{mathtools}
\usepackage{ marvosym }
\usepackage{tikz-cd}

\makeatletter
\begin{document}
\title{On the Existence and Uniqueness of Poincar\'e Maps\\ for Systems with Impulse Effects}

\author{Jacob R. Goodman 
        and~Leonardo J. Colombo% <-this % stops a space
\thanks{ J. R. Goodman is with Department of Mathematics, University of Michigan, 530 Church St. Ann Arbor, 48109, Michigan, USA.
        {\tt\small jdkgbmx@umich.edu}}% <-this % stops a space
% <-this % stops a space
\thanks{L. J. Colombo is with Instituto de Ciencias Matem\'aticas
(CSIC-UAM-UC3M-UCM), Calle Nicol\'as Cabrera 13-15, 28049, Madrid, Spain. {\tt\small leo.colombo@icmat.es}}}% <-this % stops a space
\maketitle

\begin{abstract}
The Poincar\'e map is widely used to study the qualitative behavior of dynamical systems. For instance, it can be used to describe the existence of periodic solutions. The Poincar\'e map for dynamical systems with impulse effects was introduced in the last decade and mainly employed to study the existence of limit cycles (periodic gaits) for the locomotion of bipedal robots. We investigate sufficient conditions for the existence and uniqueness of Poincar\'e maps for dynamical systems with impulse effects evolving on a differentiable manifold. We apply the results to show the existence and uniqueness of Poincar\'e maps for systems with multiple domains.
\end{abstract}


% \fi
\IEEEpeerreviewmaketitle

\section{Introduction}

Hybrid systems are non-smooth dynamical systems which exhibit a combination of both continuous and discrete dynamics. In particular, the flow evolves continuously on a state space, and a discrete transition occurs when the
flow reaches a co-dimension one hypersurface of the state space transverse to the flow \cite{brogliato}, \cite{teelSurvey}, \cite{liberzon}, \cite{van}. Due to advances in control systems, modeling, and analysis of switching and robotic systems \cite{sanfelice}, \cite{tomlin}, \cite{Varaiya}, \cite{Ye}, there has been an increased interest in studying the existence and stability of limit cycles in hybrid systems. To this end, the Poincar\'e map has become an indispensable tool \cite{revzen}, \cite{PBpaper}, \cite{de}. 

Systems with impulse effects are a class of hybrid systems with continuous dynamics typically given by a mechanical system and where the transition between the continuous and discrete behavior is determined by an impulsive (inelastic) impact. This gives rise to a discontinuity in the velocity of the system, while the trajectory is either left or right continuous. This class of hybrid system is also referred to as a \textit{simple hybrid system} \cite{SHS}, \cite{SIEs}, \cite{amesthesis}, \cite{haddad}, \cite{Biped-book}.%Such a class of hybrid systems possesses trajectories which are (left or right) continuous but there is a discontinuity in the velocity at the impact time.

\begin{figure}[h!] 
\begin{tikzpicture}[scale=0.8]
%\draw[very thick] (4,0) -- (7,0)--(8.5,2)--(5.5,2)--(4,0);
\draw[very thick] (3.8,0) .. controls (3.7,0.8) and (4.9,1.8) .. (5.7,2);%izquiera
\draw[very thick] (3.8,0) .. controls (3.8,-0.3) and (6,0.3) .. (7.5,0);%abajo
%\draw[very thick] (7.5,0) .. controls (7.2,0.5) and (7.8,1.9) .. (8,2);%derecha
%\draw[very thick] (5.7,2) .. controls (3.5,2) and (8.3,2.2) .. (8,2);%arriba
\draw [orange]  (5.75,0.9) circle (20.5pt);
\filldraw [blue]  (5.75,0.9) circle (1.5pt) node[below]{$m^{+}$};
%\draw (6.3,0.55) node[right]{$\widetilde{W}'_{0}$};
\draw (5.5,0) node[below]{$\Delta(S)$};
\draw[very thick] (8.8,0) .. controls (8.7,0.8) and (9.2,1.8) .. (10,2);%izquiera
\draw[very thick] (8.8,0) .. controls (8.8,-0.3) and (9.8,0.3) .. (12,0);%abajo
%\draw[very thick] (10,0) .. controls (9.7,0.5) and (11,1.9) .. (12,2);%derecha
%\draw [orange]  (10.7,0.5) circle (9.5pt);
\draw [orange]  (10.5,1.4) circle (13pt);
%\draw (11.1,0.65) node[right]{$W_2'$};
\filldraw [blue]  (9.75,0.6) circle (1.5pt);
\filldraw [red] (10.3,1.4) circle(1.5pt); 
\filldraw [red](10.3,1.3)node[right]{$m^{-}$};
%\draw (10.7,1.55) node[right]{$W_0'$};
\draw (10.5,-0.4) node[right]{$S$};
\draw [->]  (5.75,0.9) .. controls (7,0.1) and (7,0.1) .. (9.75,0.6);
\draw [dashed, <-] (5.75,0.9) .. controls (7,2) and (7,2) .. (10.3,1.4);
\draw (8.5,2.25) node [below]{$\Delta$};
\draw (8.2,0.3) node [below]{$F_{\delta}$};
\draw [blue] (9.8,0.6) node[right]{$\Theta(m^{-})$};
\end{tikzpicture}\caption{Illustration of the geometric setup of the Poincar\'e map for systems with impulse effects. $m^-$ and $m^+$ denote the left-limit and right-limit, respectively, of the trajectory as it intersects $S$ . Circles denote open sets around the point, and $F_{\delta}$ the continuous-time flow after a time $\delta$.}\label{figure}
\end{figure}
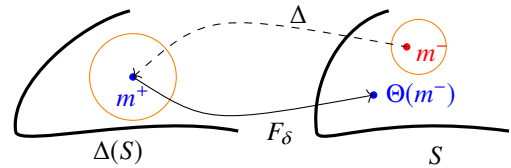

As with smooth dynamical systems, the Poincar\'e map for systems with impulse effects requires the construction of a hypersurface transversal to the periodic orbit. The return map $\Delta$ takes place on the guard $S$, which determines when the states of the dynamical system are to be reset, and provide the natural choice for the transversal hypersurface. Hence, the Poincar\'e
return map $\Theta$ for systems with impulse effects is defined on a subset of the guard which
induces a discrete-time map from this subset onto $\Delta(S)$. This map executes the trajectory of the system from a point on the guard to its next corresponding intersection with the guard. The time $\delta$ when the flow intersects the guard is called impact time. Figure \ref{figure} illustrates the situation.

The Poincar\'e map for systems with impulse effects have been introduced in \cite{grizzle} (see also \cite{Biped-book}). It was mainly employed in the search of periodic gaits (limit cycles) of bipedal robots, together with the use of several methods such as geometric abelian Routh reduction, hybrid zero dynamics and virtual constraints, hybrid Hamiltonian  systems, symmetries, etc  \cite{quasi}, \cite{fe}, \cite {grizzle2}, \cite{gregg}, \cite{spring}, \cite{hamed}, \cite{KH1}, \cite{KH2}, \cite{KH3},  \cite{ames2018}, \cite{veer1}, \cite{veer2}, \cite{veer4}. These methods permit one to gain extra information regarding the behavior of the system, which provide advantages for the construction of the Poincar\'e map.

The goal of this paper is to augment the method of Poincar\'e maps for nonlinear systems with impulse effects evolving on a differentiable manifold, and state sufficient conditions that ensure the existence and uniqueness of such a map. The problem of existence and uniqueness was not addressed even for systems evolving on Euclidean spaces. The approach provided in this work directly covers such a situation, and fixes a gap in the literature about the formal construction of this map, sufficient conditions for its existence (even for non-integrable systems) and its uniqueness. The proof of the main result is based on the sketch of proof for classical (non-hybrid) dynamical systems given in \cite{AbMarsden} (Theorem 7.1, Ch.7, pp. 521).

In most circumstances, it will be possible (and required in numerical simulatios) to work within chart maps on the manifold which defines the state space of the system, and to study the dynamics on some $\mathbb{R}^n$ locally diffeomorphic to the manifold. However, in this paper, we will adopt an intrinsic and coordinate free interpretation of the Poincar\'e map by constructing the map on the manifold. While this indeed comes with a deeper level of abstraction, it also lends itself to conceptual clarity in the proof and a more general interpretation of the map. In particular, most of the results in the literature which make use of the Poincar\'e map define the state space as $\mathbb{R}^{n}$, but these constructions do not include other important situations in engineering applications derived directly from the explicit construction of the map provided by the existence and uniqueness theorem.  For instance, when dealing with the problem of uniqueness (a natural question along with existence, and of great importance in terms of stability analysis) it is highly advantageous to adopt this intrinsic approach because the choices in atlas and charts are not unique. This leads to a variability in the representation of the Poincar\'e map which does not occur in our approach, and is not conducive to the question of uniqueness and stability. 

On the other hand, our approach does directly imply all of the desired results within charts on the manifold, so it comes at no cost for simulation purposes.  For continuous-time dynamical systems, different Poincar\'e maps on the same system have the same stability results, and we provide sufficient conditions under which we can change the guard and reset and preserve stability. The existence theorem provides a recipe for constructing the Poincar\'e map, which is useful in engineering purposes for integrable systems. In particular, the proof shows the explicit construction of the Poincar\'e map and differentiating this representation gives rise to a practical use for finding eigenvalues in order to conduct the stability analysis for periodic orbits.

%Nevertheless, in the various engineering situations covered in most of the previous works, the search for periodic solutions is done by running a numerical integration routine in local coordinates. Such local coordinates are coordinates on a particular chart of the manifold, configuration space of the system, which is locally diffeomorphic to some $\mathcal{R}^n$. But what about if the impact takes place on the intersection of two charts? Imagine the situation when...}

% \textcolor{blue}{such as Lie group configuration spaces like the special orthogonal group or the special Euclidean group. We believe that the proposed results of this paper may find use, for instance, in the screw motion of a robotic hand hitting a wall} \cite{gracioso}, \cite{lynch}. The proof of the main result is based on the sketch of proof for classical (non-hybrid) dynamical systems given in \cite{AbMarsden} (Theorem 7.1, Chapter 7, pp. 521).

The paper is organized as follows: Section $2$ introduces the class of hybrid systems we will study in this work (i.e., nonlinear systems with impulsive effects). Section $3$ states and proves the main results of this work. Finally, Section $4$ applies the result  of Section $3$ to systems with impulse effects and with multiple domains.

%%------------------------------------------------------------------ Preliminaries -----------------------
%%------------------------------------------------------------------------------------------------------------
\section{ Dynamical systems with impulse effects}
%and possibly, giving rise to a discontinuous flow
A \textit{dynamical system with impulse effects} (SIEs) is a class of hybrid dynamical system (HDS) that exhibits both discrete and continuous behaviors. The transition from one to the other is determined by the time when the continuous-time flow reaches a co-dimension one submanifold of the state space, reinitializing the flow for the ODE which specifies the continuous-time dynamics. This class of dynamical systems are characterized by a 4-tuple $\mathscr{H} = (M, S, X, \Delta)$, where:
\begin{enumerate}
\item $M$ is a differentiable manifold called  the \textit{domain},
\item $S\subset M$ is an embedded co-dimension one submanifold of $M$, called the \textit{guard}, or the \textit{switching surface}, %It is here that discrete behavior may occur.
\item $X$ is a smooth vector field on $M$ with flow $F: \mathscr{D} \to M$ where $\mathscr{D}$ is an open set in $M \times \mathbb{R}$, %[appendix lemma 4]
\item $\Delta: S \to M$ is a $C^1$ function, called the \textit{reset} or \textit{impact} map, which re-initializes the trajectories that cross $S$.
\end{enumerate}
%(more compactly, a \textit{section} of $M$) : No, I don't think iyou are using the correct name.

The pair $(M, X)$ describes the continuous-time dynamics of $\mathscr{H}$, whereas $(S, \Delta)$ defines the discrete-time dynamics. The underlying \textit{dynamical system with impulse effects} is given by:

\[\Sigma_{\mathscr{H}}: \begin{cases} 
      \dot{m} = X(m) & \ \text{ if } \ m\notin S \\
      m^+ = \Delta(m^-) & \ \text{ if } \ m^- \in S.
   \end{cases}
\]%(and with the class of systems we will continue to work with in this paper)
%We define the \textit{Hybrid Flow} of $\Sigma_{\mathscr{H}}$ by $F^\mathscr{H} = F \circ \Delta_M$. The integral curve $F^\mathscr{H}_m$ is periodic if there exists some $(m, \tau) \in M \times \mathbb{R}$ such that $F^\mathscr{H}_m(t + \tau) = F^\mathscr{H}_m(t)$ for all $t \in \mathbb{R}$. We define the corresponding perioidic orbit as \\ $\gamma = \{ m \in M \ \vert \  m = F^\mathscr{H}_{m_0}(t) \text{ for some $t \in \mathbb{R}$ where $F^\mathscr{H}_{m_0}$ is periodic} \}$
Here $m^-$ and $m^+$ denote the left-limit and right-limit, respectively, of the trajectory as it intersects $S$ (and is correspondingly reset by $\Delta$). In general, $m^- \ne m^+$, so that there may be a point of discontinuity here. However, as in  \cite{Biped-book}, we are given the choice in deciding whether the trajectory will be left-continuous or right-continuous at this point. That is, whether $m^- \in S$ or $m^+ \in \Delta(S)$ belong to our trajectory. In this paper, we will choose the former. Note that the results that follow in this work hold regardless of this choice \cite{haddad}.  However, this means that the orbits associated with the flow of this class of HDS will not (in general) be closed.

\begin{remark}{(Zeno behavior)}\label{remark1} Consider the impact map $\Delta$ given by the identity map. When a trajectory crosses $S$, we will have $m^+ = \Delta(m^-) = m^- \in S$, so that we are again in the regime of discrete dynamics where re-initialization (to $m^-$) will occur. It is clear that this process will never terminate, so that there exists an infinite number of resets in finite amount of time. This situation generates a class of behaviors called \textit{Zeno behavior}. It is particularly problematic in applications where numerical work is used, as computation time grows infinitely large at these Zeno points. While there have been proposed models for treating Zeno behavior \cite{amesthesis}, \cite{amesgregg}, \cite{teelSurvey},  \cite{kale}, \cite{lygeros}, \cite{geometrichybrid},  \cite{zeno} we exclude it from our systems, as it is not very relevant in models for locomotion (where completely plastic impacts with no rebound are assumed). There are two primary modes through which zeno behavior can occur:

1) A trajectory is reset back onto the guard, prompting additional resets. As seen in the above example, if there is a set of points in the guard which the reset map cycles between, we can get stuck' on the guard. To exclude this type of behavior, we require that $S \cap \overline{\Delta}(S) = \emptyset$, where $\overline{\Delta}(S)$ denotes the closure as a set of $\Delta(S)$. This ensures that the trajectory will always be reset to a point with positive distance from the guard.

2) The set of times where a solution to our system reaches the guard (called the set of \textit{impact times}) has a limit point. This happens, for example, in the case of the bouncing ball with coefficient of restitution $1/2$. If $t_0$ is the time between two impacts, then the time between the next two impacts will be $t_0/2$, then $t_0/4$, and so on. In time $$T = t_0 + \frac{t_0}{2} + \frac{t_0}{4} +...+ \frac{t_0}{2^n} +... = 2t_0$$
we will have infinitely many resets in finite time. To exclude these types of situations, we require that the set of impact times be closed and discrete, as in \cite{Biped-book}, and therefore avoiding to have finite accumulation points.  The above two assumptions will be assumed throughout the remainder of the paper. 

\end{remark}

By Remark \ref{remark1}, we may extend the domain of the reset to the entire manifold without affecting the dynamics by defining a function $\Delta_M: M \to M$ by:
\[ \Delta_M(m) = \begin{cases} \Delta(m) & \text{ if } m \in S, \\ m & \text{ if } m \notin S. \end{cases} \]The map $ \Delta_M(m)$ permits
us to define the flow for the SIEs $\Sigma_{\mathscr{H}}$ as follows.

\begin{definition}
Consider the SIEs $\Sigma_{\mathscr{H}}$:

\begin{enumerate}

\item The \textit{flow} of $\Sigma_{\mathscr{H}}$ is given by $F^\mathscr{H} = \Delta_M \circ F$ and the \textit{integral curve at m} by $F^\mathscr{H}_{m} := F^\mathscr{H}(m, \cdot )$
\item The integral curve $F^\mathscr{H}_m$ is \textit{periodic} if there exists some $(m, \tau) \in \mathscr{D}$ such that $F^\mathscr{H}_m(t + \tau) = F^\mathscr{H}_m(t)$ for all $t \in \mathbb{R}$. 
\item If the integral curve $F^\mathscr{H}_{m}$ is periodic, the corresponding \textit{periodic orbit} is given by: \[ \gamma = \{  m' \in M \ \vert \  m' = F^\mathscr{H}_{m}(t), \ t \in [0, \tau]  \} \] %where $F^\mathscr{H}_{m_0}$ is periodic}
\end{enumerate}
\end{definition}
%
%
%          SECTION 3
%
%

\section{Existence and uniqueness of Poincar\'e maps for dynamical systems with impulse effects}

In this section we show the main results of the paper. However, before stating the theorem and its proof, we must introduce some necessary preliminary notions and results.

\begin{definition} A \textit{section} $S$ of $M$ is a co-dimension one submanifold of $M$. $S$ is said to be \textit{locally transverse at} $s \in S$ if $X(s) \notin T_s S$ (where $X$ is the vector field described in the definition of SIEs). If $X(s) \notin T_s S$ for all $s \in S$, then $S$ is a \textit{local transverse section}.
\end{definition}
We will say that a section S of $M$ is locally transverse with respect to $X$ at $s \in S$ if $X(s) \notin T_s S$. However, we will often drop the references to our particular vector field and manifold, as it will be understood by our problem set-up.

\begin{proposition}\label{proposition}
Let $X$ be a smooth vector field with flow $F$, and let $S$ be a section of $M$ locally transverse at $s \in S$. If $\{s\} \times I \subset \mathscr{D}$ for all $s \in S$ and for some interval $I\subset\mathbb{R}$, then for all $\lambda \in I, \ F_{\lambda}(S)$ is locally transverse at $F_{\lambda}(s)$. Moreover, if $S$ is a local transverse section, then so is $F_{\lambda}(S)$.

\end{proposition}
\textit{Proof:} 
Note that by Lemma $4$ (see Appendix) $S':=F_{\lambda}(S) $ is a section of $M$ since $S$ is a section of $M$ and $F_{\lambda}$ is a diffeomorphism. Moreover, $F_s: I \to M$ is an integral curve at $s$, and $F_{\lambda} \circ F_s$ is an integral curve at $s':=F_{\lambda}(s) $ on some open interval. Assume that there exists a curve $c$ on $S'$ at $s'$ that is tangent to $F_{\lambda} \circ F_s$ at $s'$. Since $F_{\lambda}^{-1}$ is a $C^1$ mapping from a differentiable manifold to itself, we have by Lemma $2$ (see Appendix) that $F_{\lambda}^{-1} \circ F_{\lambda} \circ F_s = F_s$ and $F_{\lambda}^{-1}\circ c$ are tangent at $s$. 

Given that $F_{\lambda}^{-1}\circ c$ is a curve on $S$, it follows that $(F_{\lambda}^{-1}\circ c)'(0) \in T_s S$ and $F_s$ is an integral curve, so that $F_s'(0) = X(s) \notin T_s S$, as $S$ is locally transverse at $s$. Hence $F_s$ and $F_{\lambda}^{-1}\circ c$ are not tangent at $s$, and by contradiction, $F_{\lambda} \circ F_s$ and $c$ are not tangent at $s'$.

Since $c$ was arbitrary, $(F_{\lambda} \circ F_s)'(0) = X(s') \notin T_{s'}S'$, so $S'$ is locally transverse at $s'$. Finally, if $S$ is a local transverse section, the argument above may be applied to each of the points in $s$, from which it follows that $F_{\lambda}(S)$ is also a local transverse section.\hfill$\square$

\subsection{Existence of Poincar\'e maps for dynamical systems with impulse effects}
Next, we proceed to state and prove the main result.
\begin{theorem}\label{existence}
Let $\mathscr{H}$ be a SIEs with $S \cap \overline{\Delta}(S) = \emptyset$ and the set of impact times closed and discrete. Suppose there exists a peridic orbit $\gamma$ of $\mathscr{H}$ such that $S$ is locally transverse at $\gamma \cap S = \{m_0 \} \in S \setminus \partial S, \ \Delta(S)$ is locally transverse at $\Delta(m_0)$, and the differential of $\Delta$ is a linear isomorphism at $m_0$.

Then, there exists a map $\Theta: W_0 \to W_1$, called the \textit{Poincar\'e map}, such that
\begin{enumerate}
\item $W_0$ and $W_1$ are open sections of $S$ containing $m_0$, and $\Theta$ is a diffeomorphism
\item There exists a $C^1$ function $\delta: W_0 \to \mathbb{R}$ such that $\Theta(w) = (F_{\delta(w)} \circ \Delta)(w)$ for all $w \in W_0$, called time-to-impact map.
\end{enumerate}
\end{theorem}
\textit{Sketch of the proof:} In the following proof, we construct the Poincar\'e map explicitly. We first define a collection of suitable domains containing either $m_0$ or $\Delta(m_0)$, and a collection of maps between them so that the hybrid flow after time $\tau$ from $S$ is a local transverse section at $m_0$ contained in a straightening chart (see Lemma $3$ in Appendix). Then we look inside of this straightening chart to construct a diffeomorphism between a neighborhood of $m_0$ on $S$ and the section. Next, we compose this diffeomorphism with the hybrid flow and show that it satisfies criterion (1) of the theorem.  Finally, we show that this map can be written in the form shown in property (2), guaranteeing that it is indeed a Poincar\'e map.

\textit{Proof:}
Since $S$ is locally transverse at $m_0$, we know that $X(m_0) \notin T_{m_0}S$. But $0 \in T_{m_0}S$ since it is a vector space, so that $X(m_0) \ne 0$. Hence, we can let $(U, \phi)$ be a straightening chart at $m_0$ with $\phi: U \to V \times I \subset \mathbb{R}^{n-1} \times \mathbb{R}$ as in Lemma 3 (see Appendix). The flow at $\Delta(m_0)$ is defined on an open interval containing $[0, \tau]$, so for all $u \in U, \ \{u\}\times [0, \tau] \subset \mathscr{D}, \ $ by shrinking $U$. %as necessary. 

Since the differential of $\Delta$ is a linear isomorphism at $m_0$, by the inverse function theorem,  $\Delta$ is a local diffeomorphism at $m_0$. Let $W_0'$ be an open subset of $S$ such that $\Delta: W_0' \to \Delta(W_0')$ is a diffeomorphism, $W_0' \subset U \cap S$, and $W_0'$ is a local transverse section at $m_0$ ($S$ is a differentiable manifold and it is locally transverse at $m_0$, so it must also be locally transverse in some neighborhood of $m_0$).

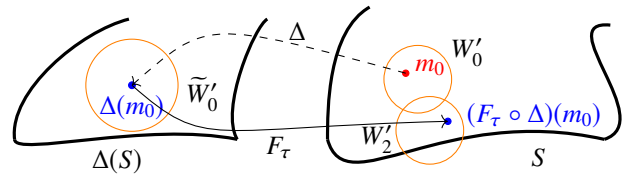
\begin{figure}[h!] 
\begin{tikzpicture}[scale=0.8]
%\draw[very thick] (4,0) -- (7,0)--(8.5,2)--(5.5,2)--(4,0);
\draw[very thick] (3.8,0) .. controls (3.7,0.8) and (4.9,1.8) .. (5.7,2);%izquiera
\draw[very thick] (3.8,0) .. controls (3.8,-0.3) and (6,0.3) .. (7.5,0);%abajo
\draw[very thick] (7.5,0) .. controls (7.2,0.5) and (7.8,1.9) .. (8,2);%derecha
%\draw[very thick] (5.7,2) .. controls (3.5,2) and (8.3,2.2) .. (8,2);%arriba
\draw [orange]  (5.75,0.9) circle (21.8pt);
\filldraw [blue]  (5.75,0.9) circle (1.6pt) node[below]{$\Delta(m_0)$};
\draw (6.5,0.75) node[right]{$\widetilde{W}'_{0}$};
\draw (5.5,0) node[below]{$\Delta(S)$};
\draw[very thick] (9,0) .. controls (9,0.5) and (9,1.8) .. (9.7,2.2);%izqierda
\draw[very thick] (9,0) .. controls (9.3,-1.1) and (10.9,0.6) .. (13.6,0);%abajo
\draw[very thick] (13.6,0) .. controls (14.5,0.3) and (12.7,2) .. (13.7,1.9);%derecha
%\draw[very thick] (9,2) .. controls (9,2.2) and (12.3,2.2) .. (13,2);%arriba
\draw [orange]  (10.7,0.15) circle (16pt);
\draw [orange]  (10.5,1) circle (16pt);
\draw (9.4,0) node[right]{$W_2'$};
\filldraw [blue]  (11,0.3) circle (1.5pt);
\filldraw [red] (10.3,1.1) circle(1.5pt); 
\filldraw [red](10.3,1.2)node[right]{$m_0$};
\draw (10.9,1.45) node[right]{$W_0'$};
\draw (12.5,0) node[below]{$S$};
\draw [->]  (5.75,0.9) .. controls (7,-0.2) and (7,0.2) .. (11,0.3);
\draw [dashed, <-] (5.75,0.9) .. controls (7,2) and (7,2) .. (10.3,1.1);
\draw (8.5,2.1) node [below]{$\Delta$};
\draw (8.2,0.2) node [below]{$F_{\tau}$};
\draw [blue] (11.16,0.4) node[right]{$(F_{\tau}\circ\Delta)(m_0)$};
\end{tikzpicture}\caption{Situation displaying the constructed sets and their relations to each other, together with the corresponding flow and reset map on these sets.}\label{figureth1}
\end{figure}

Let $V_0 = \{v \in V \ \vert \ \exists \lambda' \in I \text{ such that } (v, \lambda') \in \phi(W_0') \}$. Then $V_0 \times I \subset V \times I$ is open and contains $\phi(m_0) = 0$. Further define
\[ \begin{cases} 
      U_0 := \phi^{-1}(V_0 \times I), \\
    \widetilde{W}'_{0} := F_{\tau}^{-1}(U_0) \cap \Delta(W_0'), \\
      W_2':=F_{\tau}(\widetilde{W}'_{0}),
   \end{cases}
\]where $U_0$ is an open subset of $U$, and $W_0'$ is an open section, both containing $m_0$. Since $F_{\tau} \vert_{U}$ and $\Delta \vert_{W_0'}$ are diffeomorphisms, $\widetilde{W}'_{0}$ is an open section containing $\Delta(m_0)$. Consequently, $W_2'$ is an open section in $U_0$ containing $m_0$, and, given that $\widetilde{W}'_0$ is locally transverse at $\Delta(m_0)$, by Proposition $3.2$, $W_2'$ is locally transverse at $m_0$. Figure \ref{figureth1} illustrates the situation of the previous construction.

\begin{figure}[h!] 
\begin{tikzpicture}[scale=0.8]

\draw[very thick] (5.8,0) .. controls (0,0) and (0,0) .. (9.8,0);%abajo
\draw (10.5,0.6) node [below]{$V_0 \times \{0\}$};
\filldraw [purple] (5.655,0) circle(2pt); 
\draw (5.6,0) node [below]{$0$};
\draw [-, blue]  (1.9,-1) .. controls (1.9,1) and (9.8,-1.3) .. (9.8,2);%%%%
\draw (9.7,1.5) node [right]{$\phi(W_0')$};
\filldraw [blue] (9.5,1.08) circle(2pt); 
\draw [dashed] (9.5,-1) -- (9.5,1.08);
\draw (9.4,1.08)  node [left]{$L(m)$};
\filldraw [black] (9.5,-1) circle(2pt); 
\draw (8.8,-1) node [right]{$m$};
\draw [-, red]  (1.9,1) .. controls (1.9,-1) and (9.8,1) .. (9.8,-1);
\draw (2,1) node [right]{$\phi(W_2')$};
\draw (0,2) node [right]{$V_0\times I$};
\end{tikzpicture}\caption{Situation displaying a cross-section of the straightening chart. The point $0$ in the chart picture stands for $m_0$ (the straightening point maps $\phi(m_0) = 0$).}\label{figureth2}
\end{figure}
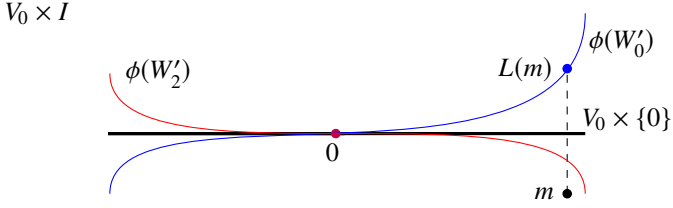

Next, we proceed to construct the Poincar\'e map. By construction of $V_0 \times I$ and the fact that $W_0'$ is a local transverse section, we have that for any $(v, \lambda) \in V_0 \times I$, there exists a unique point $T(v, \lambda) \in \mathbb{R}$ such that $(v, \lambda + T(v, \lambda)) \in \phi(W_0')$. Let $T: V_0\times I \to \mathbb{R}$ be this map, and $L: V_0\times I \to \phi(W_0')$ be the map given by $L(v, \lambda) = (v, \lambda + T(v, \lambda))$. Figure \ref{figureth2} illustrates the situation.%\footnote{\textcolor{blue}{The blue and red surfaces are \textit{not} becoming vertical at the ends as it looks in the picture at a first look (it cannot happen for transverse sections)}} 

Now consider the canonical projection $\pi: V_0 \times I \to V_0 \times \{0\}$, which sets the $n^{th}$ component of its input to $0$. For some $(\alpha_1,...,\alpha_n)\in\mathbb{R}^n$, we can write the tangent plane of $\phi(W_0')$ at $\phi(m) = 0$ as
\begin{align*}
T_0(\phi(W_0')) =& \{(x_1,...,x_n)\in\mathbb{R}^n \vert \ \alpha_1 x_1 +... + \alpha_n x_n = 0 \} \\
=& \{ (x_1,..., x_{n-1}, \frac{-1}{\alpha_n}(\alpha_1 x_1 + ... + \alpha_{n-1}x_{n-1})) \mid x_j \in \mathbb{R} \},\end{align*} where $\alpha_n \ne 0$, since $W_0'$ is locally transverse at $m_0$. It is clear from this representation that $\pi\vert_{T_0(\phi(W_0'))}$ is a linear isomorphism, so that $\pi \vert_{\phi(W_0')}$ is a local diffeomorphism at $0$ by the inverse function theorem.

Since $W_2'$ is also a subset of $U_0$ that is locally transverse at $m_0$, we have that $\pi \vert_{\phi(W_2')}$ is a local diffeomorphism at $0$ using the same argument. In summary,
\begin{itemize}
\item[(i)] $\pi \vert_{\phi(W_0')}$ maps an open subset of $\phi(W_0')$, containing $0$, diffeomorphically to an open subset of $V_0 \times \{0 \}$, say $V_1$, with $0 \in V_1$.
\item[(ii)] $\pi \vert_{\phi(W_2')}$ maps an open subset of $\phi(W_2')$, containing $0$, diffeomorphically to an open subset of $V_0 \times \{0 \}$, say $V_2$, with $0 \in V_2$.
\end{itemize}

Let $W_2 := (\phi^{-1} \circ \pi \vert^{-1}_{\phi(W_2')})(V_1 \cap V_2)$. $W_2$ is an open subset of $W_2'$ containing $m_0$ and $(\pi \vert^{-1}_{\phi(W_0')} \circ \pi \vert_{\phi(W_2)})$ maps $\phi(W_2)$ diffeomorphically onto its image, say $\phi(W_1)$, where $W_1$ is an open subset of $W_0'$. 

Note that $L\vert_{\phi(W_2)} =
\pi \vert^{-1}_{\phi(W_0)} \circ \pi \vert_{\phi(W_2)}$, since both maps preserve the element of $V_0$, $v \in V_0$. Moreover, for all $v \in V_0$ there exists a unique element $\lambda \in I$ such that $(v, \lambda) \in \phi(W_0')$, as $W_0'$ is a local transverse section. Hence $L: \phi(W_2) \to \phi(W_1)$ is a diffeomorphism. This further implies that the map $T: \phi(W_2) \to \mathbb{R}$ is $C^1$, since $L$ is $C^1$ and  $L(v, \lambda) = (v, \lambda + T(v,\lambda))$ for all $(v, \lambda) \in \phi(W_2)$. 

Next, we consider the sets \[ \begin{cases} 
 \widetilde{W}_{0} = F_{\tau}^{-1}(W_2), & \ \text{ open subset of } \tilde{W_0'} \text{ containing } \Delta(m_0), \\
      W_0 = \Delta^{-1}(\widetilde{W}_{0}), & \ \text{ open subset of } W_0' \text{ containing } m_0.
   \end{cases}
\] Then the composite map $\Theta$ given by
$$W_0 \xrightarrow{\Delta} \widetilde{W}_{0} \xrightarrow{F_{\tau}} W_2 \xrightarrow{\phi} \phi(W_2) \xrightarrow{L} \phi(W_1) \xrightarrow{\phi^{-1}} W_1$$is a diffeomorphism between two open neighborhoods of $m_0$ on $S$, so that $\Theta$ satisfies condition $(1)$ of our theorem. 

To show condition $(2)$ also holds, consider the $C^1$ mapping  $\delta^\star: W_0 \to \mathbb{R}$ defined as $\delta^\star = T \circ \phi \circ F_{\tau} \circ \Delta$. Let $\pi_n: \mathbb{R}^n \to \mathbb{R}$ be the canonical projection that maps vectors in $\mathbb{R}^n$ to their $n^{th}$ component. For $w \in W_0$,
\begin{align*}
\Theta(w) & = (\phi^{-1} \circ L \circ \phi \circ F_{\tau} \circ \Delta)(w) \\
& = (\phi^{-1} \circ L)((\pi \circ \phi \circ F_{\tau} \circ \Delta)(w), \ (\pi_n\circ \phi \circ F_{\tau} \circ \Delta)(w)) \\
& = \phi^{-1}( (\pi \circ \phi \circ F_{\tau} \circ \Delta)(w), \ (\pi_n \circ \phi \circ F_{\tau} \circ \Delta)(w) \\
&\quad\qquad+ (T \circ \phi \circ F_{\tau} \circ \Delta)(w)) \\
& = \phi^{-1}( (\pi \circ \phi \circ F_{\tau} \circ \Delta)(w), \ (\pi_n \circ \phi \circ F_{\tau} \circ \Delta)(w) + \delta^\star(w)).
\end{align*} 

Furthermore, by the Straightening out Theorem (see Lemma $3$ in Appendix), $\phi^{-1}((\pi \circ \phi \circ F_{\tau} \circ \Delta)(w), -): I \to U$ is an integral curve at $\phi^{-1}((\pi \circ \phi \circ F_{\tau} \circ \Delta)(w), 0)$, and by Lemma $4$, $F_{\Delta(w)}: [0, a) \to M$ is an integral curve at $\Delta(w), \ $ with $a > \tau$.

Noting that
\begin{align*}
F_{\Delta(w)}(\tau) & = (F_{\tau} \circ \Delta)(w) \\
& = (\phi^{-1} \circ \phi \circ F_{\tau} \circ \Delta)(w) \\
& = \phi^{-1}((\pi \circ \phi \circ F_{\tau} \circ \Delta)(w), \ (\pi_n \circ \phi \circ F_{\tau} \circ \Delta)(w)),
\end{align*} by Lemma $2$ (see Appendix), $$\phi^{-1}((\pi \circ \phi \circ F_{\tau} \circ \Delta)(w), \ t + (\pi_n \circ \phi \circ F_{\tau} \circ \Delta)(w) ) = F_{\Delta(w)}(t + \tau)$$ for all $t$ such that $t + (\pi_n \circ \phi \circ F_{\tau} \circ \Delta)(w)$ is on $I$ (this is because the flow $F_{\Delta(w)}$ must be defined everywhere that the integral curve given by $\phi^{-1}$ is defined, while the converse is not necessary true).

By construction, $\delta^\star(w) + (\pi_n \circ \phi \circ F_{\tau} \circ \Delta)(w)$ is on $I$, so that 
\begin{align*}
\Theta(w) & =  \phi^{-1}((\pi \circ \phi \circ F_{\tau} \circ \Delta)(w), \  \delta(w) + (\pi_n \circ \phi \circ F_{\tau} \circ \Delta)(w) ) \\
& = F_{\Delta(w)}(\tau + \delta^\star(w)) \\
& = (F_{\tau + \delta^\star(w)} \circ \Delta)(w).
\end{align*}

Finally, defining $\delta = \delta^\star + \tau$, we have that $$\Theta(w) = (F_{\delta(w)} \circ \Delta)(w),$$ and hence, condition $(2)$ holds. \hfill$\square$

% Explain why we need S to be locally transverse at m_0 and \Delta(S) to be locally transverse at \Delta(m_0)
\vspace{.2cm}
For the purposes of demonstrating the applications of Theorem \ref{existence} we consider the following example.
\begin{example}\label{example}
Let $M = \mathbb{R}^2$, parametrized with polar coordinates. Let $S$ be the switching surface given by  
$S = \{(r,\theta)\in\mathbb{R}\times\mathbb{S}^{1}|\theta= \frac{\pi}{2}\}$, that is, a line embedded in the manifold. Consider the smooth vector field on $M$ given by 
$X(r, \theta) = (r(1-r^2), 1)$
and the impact map $\Delta: S \to \mathbb{R}^2$ given by the rotation transformation defined by $\Delta(r, \pi/2) = (r, 0)$. Hence we consider the hybrid dynamics system determined by \[ \Sigma_{\mathscr{H}}=\begin{cases}
   (\dot{r}, \ \dot{\theta})  \ \ \ \ = (r(1-r^2), \ 1) & \ \text{ if } (r, \ \theta) \notin S \\
   (r^+, \ \theta^+) = \Delta(r^-, \ \theta^-) & \ \text{ if } (r^-, \ \theta^-) \in S.
   \end{cases}
\] Denote by $\gamma$ the circle arc of radius $1$ between $0$ and $\pi/2$. $\gamma$  is a periodic orbit of the system with period $\pi/2$. Furthermore, $S$ is locally transverse at $\gamma \cap S = (1, \pi/2)$, and the differential of $\Delta$ is a linear isomorphism at $m_0=(1, \pi/2)$. By Theorem \ref{existence}, a Poincar\'e Map $\Theta$ exists and is a diffeomorphism between two open sections of $S$ containing $m_0$.

%While it is generally very difficult, if not impossible to construct a Poinca\'e Map explicity (even in continuous systems), this example was chosen because one may construct the Poincar\'e Map explicitly.

Next, consider an initial value $(r, \pi/2)$ on $S$. After a time greater than $0$, we have moved to the point $(r, 0)$, via a rotation specified by $\Delta$. %The system then exhibits continuous dynamics on the sector from $0$ to $\pi/2$. 
Since $\dot{\theta} = 1$, it follows that the trajectory will return to $S$ after time $t = \pi/2$. Integrating the $r$-component of our equation, we have that $\Theta(r)$ must satisfy:

$$\int_{r}^{\Theta(r)} \frac{dr}{r(1-r^2)} = \int_{0}^{\pi/2} dt,$$ which upon integration and solving for $\Theta(r)$ we obtain 

$$\Theta(r) = [1 + e^{-\pi}(r^{-2} - 1)]^{-1/2}.$$

Computing the derivative of $\Theta$, we get:

$$\Theta'(r) = \frac{e^{-\pi}}{r^3[e^{-\pi}(\frac{1}{r^2} - 1)+1]^{3/2}},$$ which is non-zero and continuous everywhere except for $r = 0$. Hence $\Theta$ is $C^1$ on $S$, and by the inverse function theorem, has an inverse which is $C^{1}$ in a neighborhood of $(r, \theta) = (1, \pi/2)$.  Thus, $\Theta$ is a diffeomorphism between two open neighborhoods of $(1, \pi/2)$ on $S$ by Theorem \ref{existence}.
\end{example}%(there are other singularities, but they are complex since $e^{-\pi} < 1$)
\subsection{Uniqueness of Poincar\'e maps for dynamical systems with impulse effects}
In a continuous-time dynamical system, uniqueness of Poincar\'e maps is equivalent to local conjugacy (see Theorem \ref{thcontinuous} in the Appendix). This is a desirable property, as local conjugacy preserves the eigenvalues for the Jacobian of the Poincar\'e map, for which stability analysis is concerned (see \cite{AbMarsden} Chapter $7$). If we did not have uniqueness, different Poincar\'e maps may give rise to conflicting stability results for the same periodic orbit, invalidating one of the primary applications of these functions.

Local conjugacy (and uniqueness) will be defined in a similar fashion as continuous-time systems. However, we must take caution, as SIEs and its periodic orbits are intimately connected to the chosen reset and guard, which is not the case in continuous-time systems. First, we introduce the definition of local conjugacy that we will be using throughout this section.

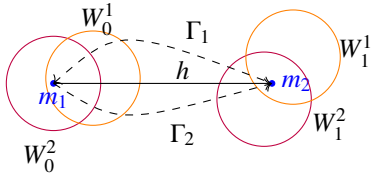
\begin{figure}[h!] \begin{center}
\begin{tikzpicture}[scale=0.7]
\draw [purple]  (5.75,0.9) circle (25.5pt);
\draw [orange]  (6.5,1) circle (25.5pt);
\filldraw [blue]  (5.75,0.9) circle (1.5pt) node[below]{$m_{1}$};
\draw (5.5,0) node[below]{$W_0^2$};
\draw (6.6,2.6) node[below]{$W_0^1$};

\filldraw [blue]  (9.9,0.9) circle (1.5pt) node[right]{$m_{2}$};
\draw [purple]  (9.75,0.6) circle (25.5pt);
\draw [orange] (10.3,1.4) circle (25.5pt); 
\draw (11.5,2) node[below]{$W_1^1$};
\draw (11,0.6) node[below]{$W_1^2$};
%\draw (10.7,1.55) node[right]{$W_0'$};

\draw [dashed, ->]  (5.75,0.9) .. controls (7,0.1) and (7,0.1) .. (9.9,0.9);
\draw [dashed, <-] (5.75,0.9) .. controls (7,2) and (7,2) .. (9.9,0.9);
\draw [ ->] (5.75,0.9) to (9.9,0.9);
\draw (8.5,2.25) node [below]{$\Gamma_1$};
\draw (8.2,0.3) node [below]{$\Gamma_2$};
\draw (8.2,0.8) node [above]{$h$};
\end{tikzpicture}\caption{Illustration of the geometric setup for locally conjugate diffeomorphisms}\label{figure5}\end{center}
\end{figure}

\begin{definition}
Two diffeomorphisms $\Gamma_1:W_0^{1}\to W_1^{1}$ and  $\Gamma_2:W_0^{2}\to W_1^{2}$ between open sets are said to be locally conjugate at the points $m^1 \in W_0^{1} \cap W_1^{1}$ and $m^2 \in W_0^{2} \cap W_1^{2}$ if there exist open sets $W^{1}\subset W_0^{1}\cap W_1^{1}$ containing $m^1$ and $W^{2}\subset W_0^{2}\cap W_1^{2}$ containing $m^2$ and a diffeomorphism $h:W^1\to W^2$ such that $\Gamma_2\circ h=h\circ\Gamma_1$. Figure \ref{figure5} illustrates the situation.
\end{definition}

In the next remark we briefly discuss the use of Poincar\'e Maps in stability, and its connection to local conjugacy.
\begin{remark}\label{rkpm}
Let $\Theta$ be a Poincar\'e map for a continuous-time dynamical system on $\mathbb{R}^n$ and denote by $m_0\in \mathbb{R}^n$ the intersection point between the periodic orbit and local transverse section $S\subset T\mathbb{R}^n$ on which the Poincar\'e map is defined (and therefore a fixed point for $\Theta$). Let $m_1\in\mathbb{R}^n$ be a small perturbation of $m_0$ such that  $m_0 + m_1$ remains in $S$, and is within the domain of $\Theta$. Then, for some $m_2\in \mathbb{R}^n$, by using a Taylor expansion, we have
\begin{align*}
m_0 + m_2 &= \Theta(m_0 + m_1) \\
&= \Theta(m_0) + D\Theta_{m_0}(m_1) + O(||m_1||^2) \\
&= m_0 + D\Theta_{m_0}(m_1) + O(||m_1||^2),
\end{align*}where $D\Theta_{m_0}$ denotes the Jacobian of $\Theta$ at $m_0$ and $D\Theta_{m_0}m_1$ is the Jacobian of $\Theta$ at $m_0$ applied to $m_1$ (which can be understood in $\mathbb{R}^n$ as the directional derivative of $\Theta$ at $m_0$ in the direction of $m_1$). Hence, $m_2 \approx  D\Theta_{m_0}(m_1)$ for sufficiently small perturbations.
 
Denoting by $\Theta^k$ the composition of $\Theta$ with itself $k$-times, it is not difficult to see that $\displaystyle{\lim_{k\to\infty}\Theta^k(m_0+ m_1)}=m_0$ if and only if all the eigenvalues for the Jacobian $D\Theta_{m_0}$ are within the init circle.

Now, in the light of the definition for local conjugacy, suppose we have a diffeomorphism $h$ with the proper domain and image so that $h \circ \Theta \circ h^{-1}$ is well-defined. Then,\begin{align*}
D(h \circ \Theta \circ h^{-1})_{h(m_0)} &= Dh_{m_0}D\Theta_{m_0}Dh^{-1}_{h(m_0)} \\
&=  Dh_{m_0}D\Theta_{m_0}(Dh)^{-1}_{m_0},
\end{align*} and so denoting by $I$ the $n\times n$ identity matrix  
\begin{align*}
&\det(D(h \circ \Theta \circ h^{-1})_{h(m_0)} - \lambda I) = \det( Dh_{m_0}D\Theta_{m_0}(Dh)^{-1}_{m_0} - \lambda I )=\\
&\det ( Dh_{m_0}( D\Theta_{m_0} - \lambda I ) (Dh)^{-1}_{m_0} )=\\
&\det ( Dh_{m_0})\det( D\Theta_{m_0} - \lambda I)\det( (Dh)^{-1}_{m_0})=\\
&\det ( Dh_{m_0})\det( D\Theta_{m_0} - \lambda I)\det( Dh_{m_0})^{-1} =\\
&\det( D\Theta_{m_0} - \lambda I).
\end{align*}
Therefore the eignevalues of $D\Theta_{m_0}$ and $D(h \circ \Theta \circ h^-1)_{h(m_0)}$ are the same.

Now, in the case that we are working on a differentiable manifold, we may choose a local representative for the Poincar\'e map and carry on with a similar analysis. The same idea holds for Poincar\'e maps on SIEs, as the map itself still remains a diffeomorphism between differentiable manifolds (in particular, of local transverse sections).
\end{remark}

Next, we provide an example where uniqueness (local conjugacy) fails for Poincar\'e maps on SIEs, demonstrating the role of the reset map in this notion.
\begin{example}
Consider Example \ref{example} but now with reset map given by $\Delta(r, \pi/2) = (e^{c(r-1)}, 0)$, with $c \in \mathbb{R}^+$. Note that $\Delta$ is $C^1$ and is a diffeomorphism between $S$ and $\Delta(S)$. Hence, by Theorem \ref{existence}, a Poincar\'e map exists.

Proceeding as before, integrating the $r$-component, we find that $$\Theta(r) = [1 + e^{-\pi}(e^{-2c(r-1)}-1)]^{-1/2},$$ so that $$\Theta'(r) = ce^{-2c(r-1)-\pi}[1+e^{-\pi}(e^{-2c(r-1)-1})]^{-3/2}.$$

As in Example \ref{example}, the circular arc of radius $1$ between $0$ and $\pi/2$ denoted by $\gamma$, is a periodic orbit for the system, and it is independent of $c$. Note that $\Theta'(1) = ce^{-\pi}$, so that systems with different values of $c$ cannot be locally conjugate by the Remark \ref{rkpm}. Moreover, note that for $c > e^\pi$, the periodic orbit $\gamma$ is unstable, and for $c < e^\pi$, $\gamma$ is stable.
\end{example}

\begin{lemma}
Let  $\Sigma_{\mathscr{H}} = (M, S, X, \Delta)$ be a SIEs satisfying the hypothesis of Theorem \ref{existence}. If $\Theta^1: W^1_0 \to W^1_1$ and $\Theta^2: W^2_0 \to W^2_1$ are two Poincar\'e Maps of $\Sigma_{\mathscr{H}}$ on $S$, then $\Theta^1 = \Theta^2$ on the intersection of their domains.
\end{lemma}\label{lemmau1}
\textit{Proof:} Here we use the same notation as the proof of Theorem \ref{existence}.  Since for all $j=0,1$ and $i=1,2$, $W^i_j$ must contain $m_0$, it is clear that the intersection $W_0 := W^1_0 \cap W^2_0$ is an open section on $S$ containing $m_0$. Let $\delta^1$ and $\delta^2$ be time-to-impact times such that $\Theta^1 = F_{\delta^1} \circ \Delta$ and $\Theta^2 = F_{\delta^2} \circ \Delta$. Note that, given $m\in W_0$ arbitrary,  $\Theta^1(m)$ and $\Theta^2(m)$ are described by the same integral curve after evolving for times $\delta^1(m)$ and $\delta^2(m)$ from the same point $\Delta(m)$. By construction and the transversality of $S$, the time-to-impact gives the smallest time at which the flow reaches $S$ from $\Delta(m)$. Since the time-to-impact map gives the smallest time after which $F_{\delta}$ reaches the guard from a point on $\Delta(W^i_0)$ they both provide "minimal times" and then they must be equal (where defined). That is, the time-to-impact map is unique. Therefore, $\delta^1 = \delta^2$ on $W_0$, hence, $\Theta^1 = \Theta^2$ on $W_0$.\hfill$\square$

%The time-to-impact map gives the smallest time after which F_{\delta} reaches the guard from a point on \Delta(W^i_0). In this sense, they both provide "minimal times", and since \delta^1 and \delta^2 are both defined on W_0, we have that they must be equal.

Lemma $3.8$ states that a Poincar\'e map within a given SIEs satisfying the hypotheses of Theorem \ref{existence} is unique (up to the choice in domain). In this sense, the natural choice for \textit{the} Poincar\'e Map would be the one with maximal domain. For the remainder of the section, we will establish a notion of uniquenss between \textit{different} SIEs. This is desireable because it allows us to change the reset map and guard of a SIEs without affecting the stability results.% There are a number of cases to address, so we will break them into multiple results rather than one large theorem.}

The next result states that, if we have two SIEs with the same underlying continuous-time dynamics and guard, and resets that are `continuations' of one another under the flow, then the Poincar\'e Maps on these SIEs are identical.

\begin{theorem}\label{uniqueness1}
Let $\Sigma_{\mathscr{H}_1} = (M, S, X, \Delta^1)$ and $\Sigma_{\mathscr{H}_2}= (M, S, X, \Delta^2)$ be two SIEs satisfying the hypotheses of Theorem \ref{existence} with $\Theta^1: W_0^1 \to W_1^1$ and $\Theta^2: W_0^2 \to W_1^2$. Suppose that there exists an open neighborhood  $W \subset W_0^1 \cap W_0^2$ of $m_0$ and a $C^1$-function $T: \Delta^1(W) \to \mathbb{R}$ such that $F_T \circ \Delta^1 = \Delta^2$ on some open set $W$. Then $\Theta^1 = \Theta^2$ on $W$.
\end{theorem}

\textit{Proof:} By Theorem 3.3, for some functions $\delta^1$ and $\delta^2$, we have $\Theta^1 = F_{\delta^1} \circ \Delta^1$ and
\begin{align*}
\Theta^2 &= F_{\delta^2} \circ \Delta^2 \\
&= F_{\delta^2} \circ F_T \circ \Delta^1 \\
&= F_{\delta^2 + T} \circ \Delta^1
\end{align*} on $W$. By the same argument given in Lemma 3.8 (i.e., uniqueness of time-to-impact function), we have that $\delta^2 + T = \delta^1$ on $W$. Hence, $\Theta^1 = \Theta^2$ on $W$.\hfill$\square$

% \textcolor{red}{The previous result states that, if we have two SIEs with the same underlying continuous-time dynamics and guard, and resets that are 'extensions' of one another under the flow, then the Poincar\'e Maps on these SIEs are identical.}
%Note that, by Lemma 4.4 in the Appendix, we need only have that

The next result states that, given two SIEs with the same underlying continuous-time dynamics, the Poincar\'e maps for these systems will be locally conjugate to each other if the guards and resets are continuations of each other under the flow. 

\begin{theorem}\label{uniqueness2}
Let $\Sigma_{\mathscr{H}_1} = (M, S^1, X, \Delta^1)$ and $\Sigma_{\mathscr{H}_2}= (M, S^2, X, \Delta^2)$ be two SIEs satisfying the hypotheses of Theorem \ref{existence} with $\Theta^1: W_0^1 \to W_1^1$ and $\Theta^2: W_0^2 \to W_1^2$. Suppose that there exist open sections $W^1 \subset W_0^1$ containing $m_0^1$ and $W^2 \subset W_0^2$ containing $m_0^2$  and a $C^1$-function $T: W_1 \to \mathbb{R}$ such that $F_T(W^1) = W^2$ and $\Delta^2 \circ F_T = \Delta^1$ on $W^1$. Then $\Theta^1$ is locally conjugate to $\Theta^2$ with conjugate function $h = F_T$.
\end{theorem}

\textit{Proof:} Let $\Theta^1 = F_{\delta^1} \circ \Delta^1$ and $\Theta^2 = F_{\delta^2} \circ \Delta^2$. Then,
\begin{align*}
F^{-1}_T \circ \Theta^2 \circ F_T &= F_{-T} \circ \Theta^2 \circ F_T \\
&= F_{-T} \circ (F_{\delta^2} \circ \Delta^2) \circ F_T \\
&=( F_{-T} \circ F_{\delta^2}) \circ (\Delta^2 \circ F_T) \\
&= F_{\delta^2 - T} \circ \Delta^1
\end{align*}
The result again follows by the uniqueness of the time-to-impact map. \hfill$\square$

\begin{remark}By Lemma $5$ (see Appendix), the condition that $F_{\lambda}(m_0^1) = m_0^2$ gives us sets $W^1$ and $W^2$ and uniquely defines $T$ (up to choices in $W^1$ and $W^2$) such that $F_{T}(W^1) = W^2$. Hence, we can relax the conditions slightly by demanding $F_{\lambda}(m_0^1) = m_0^2$ and checking that for some sets $W^1$ and $W^2$ and the map $T$ they induce, we have $\Delta^2 \circ F_T = \Delta^1$ on $\Delta^1$.\end{remark}

\begin{remark}
Note that in both of the previous Theorems, $\Sigma_{\mathscr{H}_1}$ and $\Sigma_{\mathscr{H}_2}$ need not have the same periodic orbit, but if they do not, one of the orbits is a continuation of the other under the flow. More precisely, let $\gamma_1$ and $\gamma_2$ be periodic orbits through the points $m_0^1$ and $m_0^2$ on $S^1$ and $S^2$, respectively. Either $\gamma_1$ or $\gamma_2$ will contain both $m_0^1$ and $m_0^2$. Without loss of generality, assume that it is $\gamma_1$. Then $\gamma_2 \subset \gamma_1$ and $$\gamma_1 \setminus \gamma_2 = \{m \in M: \ m = F_t(m_0^2) \ \text{ for } \ 0 < t < T(m_0^2) \}.$$
From this perspective, the result is intuitive. Though $\Sigma_1$ and $\Sigma_2$ are different systems describing different periodic orbits, they share continuous dynamics and have discrete dynamics that are related though the continuous components.
\end{remark}

\section{Applications to systems with Impulse Effects and Multiple Domains}
In this section we employ Theorems \ref{existence}, \ref{uniqueness1} and \ref{uniqueness2} for SIEs and with multiple domains.

The notion of a SIEs can be naturally extended to include multiple domains and resets.
\subsubsection*{Definition:}
A $k$-domain SIEs is a tuple
$$\mathscr{H} = (\Gamma, M, S, \Delta, X)$$
where
\begin{itemize}
\item[(i)] $\Gamma = (\mathcal{V}, \mathcal{E})$ is a directed graph such that $\mathcal{V} = \{q_1,..., q_k\}$ is a set of $k$ vertices, and $\mathcal{E} \subset Q \times Q$ is the  set of edges. We further define the maps $sor$ and $tar$, which return the source and target of the edge. More precisely, if $e_{ij} = (q_i, q_j)$, then $sor(e_{ij}) = q_i$ and $tar(e_{ij}) = q_j$.
\item[(ii)] $M = \{M_q\}_{q \in \mathcal{V}}$ is a collection of differentiable manifolds.
\item[(iii)] $S = \{S_e\}_{e \in \mathcal{E}}$ is a collection of guards, where $S_e$ is assumed to be an embedded open section of $M_{sor(e)}$.
\item[(iv)] $\Delta = \{\Delta_e\}_{e \in \mathcal{E}}$ is a collection of reset maps, which are $C^1$ mappings where $\Delta_e: S_e \to M_{tar(e)}$.
\item[(v)] $X = \{X_q\}_{q \in \mathcal{V}}$ is a collection of smooth vector fields.
\end{itemize}
We further define
\begin{itemize}
\item[(vi)] $\Lambda = \{0,1,2,...\} \subset \mathbb{N}$, an indexing set
\item[(vii)] $\rho: \Lambda \to \mathcal{V}$ a map recursively defined by $e_{\rho(i)} = (\rho(i), \rho(i+1))$.
\end{itemize}

The underlying dynamical system with impulse effects is then defined by
\[ \begin{cases} 
      \dot{x} = X_{i}(x) & \ \text{ if } \ x \in M_{i} \text{ and } x \notin S_i \\
      x^+ = \Delta_i(x^-) & \ \text{ if } \ x^- \in S_i,
   \end{cases}
\] where it is understood that $X_i = X_{\rho(i)}$ and similarly for $M_i$, and $S_i = S_{e_{\rho(i)}}$ and similarly for $\Delta_i$.

As before, we will assume the flow to be left continuous and will exclude Zeno behavior from this system by imposing the constraints that $S_i \cap \overline{\Delta_i}(S_i) = \emptyset$ for all $i\in \Lambda$ and the set of impact times for any solution is closed and discrete.

A periodic orbit is defined analgously for a $k$-domain SIEs as a single domain SIEs. 
\begin{theorem}\label{th4.1}
Let $\gamma$ be a periodic orbit of the $k$-domain SIEs $\mathscr{H} = (\Gamma, M, S, \Delta, X)$, and $\Lambda, \rho$ be defined by conditions (vi) and (vii) respectively. Assume that $S_i$ is locally transverse at $\gamma \cap S_i = m_i^\star$. If the differential of $\Delta_i$ is a linear isomorphism at $m_i^\star$ for all $i$, then there exists a Poincar\'e map $\Theta: W^0_0 \to W^1_0$ such that:
\begin{enumerate}
\item $W^0_0$ and $W^1_0$ are open subsections of $S_0$ containing $m_0^\star$, and $\Theta$ is a diffeomorphism between them.
\item There exists a collection of $C^1$ time-to-impact functions $\delta_i: W^0_i \to \mathbb{R}$ such that 
$$\Theta(w) = F_0^{\delta_{N-1}} \circ \Delta_{N-1} \circ F_{N-1}^{\delta_{N-2}} \circ\Delta_{N-2}\circ \ldots \circ F_1^{\delta_0} \circ \Delta_0(w)$$ where $W^0_i$ are open subsections of $S_i$ containing $m_i^0$, $w\in W_0^{0}$ and $F_i^t$ is the flow of the vector field $X_i$ after time $t$.
\end{enumerate}
\end{theorem}
% Proof

\textit{Proof:} Using the same argument as in the proof of Theorem \ref{existence}, there exists an open subset of $S_0$ containing $m_0^\ast$, say $W_0^0$, an open subset of $S_1$ containing $m_1^\ast$, say $W_1^1$, and a $C^1$-function $\delta_0: W_0^0 \to \mathbb{R}$ such that $F_1^{\delta_0} \circ \Delta_0: W_0^0 \to W_1^1$ is a diffeomorphism.

We may again apply the argument to find an open subset of $S_1$ containing $m_1^\ast$, say $W_1^0$, an open subset of $S_2$ containing $m_2^\ast$, say $W_2^{1'}$, and a $C^1$-function $\delta_1: W_1^0 \to \mathbb{R}$ such that $F_2^{\delta_1} \circ \Delta_1: W_1^0 \to W_2^{1'}$ is a diffeomorphism. Now, by shrinking $W_0^0$ as necessary, we may shrink $W_1^1$ so that it is contained in $W_1^0$. Letting $W_2^1 := (F_2^{\delta_1} \circ \Delta_1)(W_1^1)$, we get that  $$F_2^{\delta_1} \circ \Delta_1 \circ F_1^{\delta_0} \circ \Delta_0 : W_0^0 \to W_2^1$$ is a diffeomorphism. Continuing this argument until we return to $m_0^\ast$ (which must eventually occur because $\gamma$ is periodic) and relabeling as necessary, we obtain the desired result.\hfill$\square$

\begin{remark}
The notion of uniqueness described by Lemma 3.8, Theorem 3.9, and Theorem 3.10 naturally extend to SIEs with multiple domains. Let the $\mathscr{H}$ be as in the hypotheses of Theorem \ref{th4.1}, and $\Theta$ the corresponding Poincar\'e map. We write $\Theta$ as
$\Theta = \Theta_N \circ \Theta_{N-1} \circ \ldots \circ \Theta_1$, where $\Theta_i = F_i^{\delta_{i-1}} \circ \Delta_{i-1}$ and $F_N := F_0$. Note that the domain for $F_i$ is different than the domain for $\delta_{i-1}$ and $\Delta_{i-1}$, so that there are two domains needed to specify $\Theta_i$. Moreover, none of the maps $\Theta_i$ are Poincar\'e maps in their own right, as the restriction of the flow to its domains cannot be a periodic orbit. However, analogous to Lemma 5 in the Appendix, we may think of these maps in the same regard as ($2$-domain) Poincar\'e maps between two sections, and the corresponding analysis on these maps is identical to that of \textit{actual} Poincar\'e maps. Now, let $\Theta^\ast$ be a Poincar\'e Map on $\mathscr{H}^\ast := (\Gamma, M, S^\ast, \Delta^\ast, X)$ with $\Lambda^\ast=\Lambda$ and $\rho^\ast = \rho$, such that
$$\Theta^\ast = \Theta_N \circ \Theta_{N-1} \circ \ldots \circ \Theta^\ast_i \circ \ldots\ \circ \Theta_1.$$
That is, the Poincar\'e maps $\Theta$ and $\Theta^\ast$ are identical aside from one reset and guard. Restricting our attention to the (common) domains of $\Theta_i$ and $\Theta_i^\ast$, we may reduce the situation of uniquenss to one with identical analysis to uniqueness on a single domain SIEs. Following this, one may address inductively the problem when \textit{all} resets and guards are different, i.e.
$\Theta^\ast = \Theta_N^\ast \circ \Theta_{N-1}^\ast \circ \ldots \circ \Theta_1^\ast$.
\end{remark}

\section*{Appendix}\label{appendix}
Because HDS, and in particular SIEs, are a mixture of continuous and discrete
dynamics in this appendix we review some properties of vector fields used in the work for both
for continuous  dynamical systems. Lemmas $1$-$4$ below, and their proofs, can be found in \cite{AbMarsden} Chapter $2$, while theorem $4.3$ can be found in Chapter $7$.

Let $M$ and $N$ be differentiable manifolds and $X$ a vector field on $M$.

\textit{Lemma 1:} Suppose $c_1$ and $c_2$ are integral curves of $X$ at $m \in M$. Then $c_1 = c_2$ on the intersection of their domains.
\vspace{0.1cm}

\textit{Lemma 2:} Suppose $c_1$ and $c_2$ are curves at $m \in M$ which are tangent at $m$. Let $f: M \to N$ be $C^1$. Then $f \circ c_1$ and $f \circ c_2$ are tangent at $f(m)\in N$
\vspace{0.1cm}

\textit{Lemma 3:} (Straightening out theorem).
Suppose that for $m \in M, \ X(m) \ne 0$. Then there exists a local chart $(U, \phi)$ with $m \in U$ and $\phi(m) = 0$ called  \textit{straightening chart} at $m$ such that
\begin{enumerate}
\item $\phi(U) = V \times I \subset \mathbb{R}^{n-1} \times \mathbb{R}, \ V$ open, and $I = (-a, a)$ with $a>0$
\item $\phi^{-1}(v, -): I \to M$ is an integral curve of $X$ at $\phi^{-1}(v,0)$ for all $v \in V$
\item $T \phi \circ X \vert_U = \hat{e}_n \ $ (the image of integral curves in $U$ under $\phi$ are straight lines passing orthogonally to the fibers)
\end{enumerate}

\textit{Lemma 4:} Suppose $X$ is smooth and let $\mathscr{D}$ denote the set of $(m, \lambda) \in M \times \mathbb{R}$ such that there exists an integral curve of $X$ at $m$ whose domain contains $\lambda$. Then:
\begin{enumerate}
\item $M \times \{0\} \subset \mathscr{D}$
\item $\mathscr{D}$ is open in $M \times \mathbb{R}$
\item There exists a unique map $F: \mathscr{D} \to M$ such that the mapping $F_m: I \to M$ defined by $F_m(t) = F(m,t)$ is an integral curve for all $m \in M$ and some open interval $I$.
\item If $U$ is a subset of $M$ such that $\{u\}\times J \subset \mathscr{D}$ for some open interval $J$, then for all $\lambda \in J$, the mapping $F_{\lambda}$ restricted to $U$ is a diffeomorphism onto its image.
\end{enumerate}

\begin{theorem}\label{thcontinuous}{(Existence and uniqueness for continuous-time Poincar\'e map)}
Let $X$ be a smooth vector field on a differentiable manifold $M$ with integral $F$, $\gamma$ a closed orbit of $X$ with period $\tau$, and $S$ a local transversal section of $X$ at $m \in \gamma$. Then, 
\begin{enumerate}
\item there exists a Poincar\'e Map $\Theta: W_0 \to W_1$ of $\gamma$, where
\begin{itemize}
\item[(i)] $W_0, W_1 \subset S$ are open neighborhoods of $m \in S$, and $\Theta$ is a diffeomorphism
\item[(ii)] There is a $C^1$ function $\delta: W_0 \to \mathbb{R}$ such that for all $s \in W_0$, $\Theta(s) = F(s, \delta(s))$
\end{itemize}
\item If $\Theta:W_0 \to W_1$ is a Poincar\'e map of $\gamma$ (in a locally transverse section $S$ at $m\in\gamma$) and $\Theta'$ also (in $S'$ at $m' \in \gamma$), then $\Theta$ and $\Theta'$ are locally conjugate. That is, there are open neighborhoods $W_2$ of $m\in S$, $W_2'$ of $m' \in S'$, and a diffeomorphism $H: W_2 \to W_2'$ such that $W_2 \subset W_0 \cap W_1$, $W_2' \subset W_0' \cap W_1'$ and $\Theta' \circ H = H \circ \Theta$.
\end{enumerate}
\end{theorem}

The following Lemma states that if we introduce a new section to the continuous-time dynamics which intersects the periodic orbit transversely, we can construct a Poincar\'e map between two local transversal sections $S$ and $S^\ast$.

\textit{Lemma 5:}
Let $M, X, S, F$ be as in Theorem \ref{thcontinuous} and $S^\ast$ a local transversal  section at $m^\ast := F_{\lambda}(m) \in S^\ast$. There exist open neighborhood $W$ of $S$ containing $m$ and $W^\ast$ of $S^\ast$ containing $m^\ast$ and a unique $C^1$-function $T: W \to \mathbb{R}$ such that $F_T$ is a diffeomorphism from $W$ onto $W^\ast$.

\section*{Acknowledgment}

J. Goodman was supported by “Severo Ochoa Programme for Centres of Excellence” (SEV-2015-0554) through a JAE Intro Fellowship. L. Colombo was supported by MINECO (Spain) grant MTM2016-76702-P and SEV-2015-0554.
We thank Prof. A. Bloch and Dr. D. Mart\'in de Diego for valuable feedback and discussion on the paper. The authors are indebted with the reviewers for their recommendations that helped to improve the quality, clarity and exposition of this work.
\ifCLASSOPTIONcaptionsoff
  \newpage
\fi

\end{document}